\documentclass[jcp, reprint,twocolumn,noshowpacs]{revtex4}
\usepackage{epstopdf}
\usepackage{amssymb}
\usepackage{amsmath}
\usepackage{amsfonts}
\usepackage{graphicx}
\begin{document}
\def\br{{\bf r}}
\def\vcmi{v_{cm,i}}

\title{Multiscale Modeling of Binary Polymer Mixtures: Scale Bridging in the Athermal and Thermal Regime}
\author{J. McCarty}
\author{M.G. Guenza\footnote{Author to whom correspondence should be addressed. Electronic mail: mguenza@uoregon.edu}}
\affiliation{Department of Chemistry and Institute of Theoretical Science, University of Oregon, Eugene, Oregon 97403}
\date{\today}

\begin{abstract}
Obtaining a rigorous and reliable method for linking computer simulations of polymer blends and composites at different length scales of interest is a highly desirable goal in soft matter physics. In this paper a multiscale modeling procedure is presented for the efficient calculation of the static structural properties of binary homopolymer blends. The procedure combines computer simulations of polymer chains on two different length scales, using a united atom representation for the finer structure and a highly coarse-grained approach on the meso-scale, where chains are represented as soft colloidal particles interacting through an effective potential. A method for combining the structural information by inverse mapping is discussed, allowing for the efficient calculation of partial correlation functions, which are compared with results from full united atom simulations. The structure of several polymer mixtures is obtained in an efficient manner for several mixtures in the homogeneous region of the phase diagram. The method is then extended to incorporate thermal fluctuations through an effective $\chi$ parameter. Since the approach is analytical, it is fully transferable to numerous systems.
\end{abstract}
\maketitle
\section{Introduction}
Mixtures of polymers with different compositions are of great interest for many technological and industrial applications. For example, multicomponent polymer mixtures facilitate the custom tailoring of desired physical properties, leading to the creation of new materials with enhanced performance.\cite{Sperling} However, since new materials are typically processed in their liquid phase, the ability to predict the mixing behavior of various polymer composites is a desirable computational task. Because of the underlying phase diagram, polymer mixtures can phase separate at specific thermodynamic conditions of temperature and polymer volume fraction. The structure and dynamics of each component strongly depends on the region of the phase diagram that is explored, and different length scales characterize the thermodynamic and dynamical behavior of the mixture, depending on how far the system is from its phase separation.

While computer simulations are useful in elucidating the structural and dynamic complexity of these systems, they are limited in the extent of information that can be collected by the precision of the calculation, which degrades with the number of simulation steps, and by the power of the available computational machines.\cite{simul} Already for polymer liquids it is not possible to collect all the relevant information from a single simulation when long polymers are involved, as information spreads over many order of magnitude in time and length, going from the monomer length scale, $l$,  to the overall polymer dimension of the radius-of-gyration, $R_g=\sqrt{N} l/6$. The latter scales as the square root of the degree of polymerization, $N$, which can be fairly large for polymers of interest.\cite{binderbook} The added complexity of dealing with polymer mixtures arises from the presence of a new length scale of concentration fluctuations, $\xi_\phi$, which diverges as the system approaches its spinodal curve. The big challenge for such systems becomes simulating the mixtures not only at different temperatures and compositions, but also with a box size that increases with the scale of concentration fluctuations. Furthermore, the equilibration step in molecular dynamics (MD) simulations becomes extremely slow, as the time scale in which polymer mixtures phase separate diverges as the temperature approaches the critical point. Finally the timescale of relaxation of the mixture depends on the proximity of the system to the glass transition of its components. In conclusion, classical MD simulations of polymer mixtures have been limited so far to high temperatures close to the athermal limit. In that region the thermodynamic behavior is mainly guided by entropy and local packing\cite{binderbook}, while enthalpic effects may be accounted for within a high temperature perturbative framework.\cite{JARAMILLO, HEINE}

In order to extend the ability of MD simulations to make quantitative predictions about polymer blends at temperatures approaching the spinodal, information is needed on multiple length-scales of interest. As the system approaches its phase transition and the largest length scale of interest diverges, one might be tempted to neglect the details of the local scale; however, such a procedure risks losing pertinent information, since the monomeric structure of a blend's component largely determines its glass transition temperature, as well as the shape of its phase diagram. Nonetheless, capturing the global structure is equally as important since an accurate determination of the thermodynamics of the system requires obtaining the total distribution of particles in the ensemble. For example, while system-specific local information is contained within the correlation hole region of the pair distribution function, g(r), the global structure is still required, as truncation of g(r) will prevent accurate Fourier transform to the structure factor where the low k limit is important.\cite{ALLEN} Thus, both local and global scale information need to be accounted for to achieve a complete thermodynamic and dynamic description of the mixture. 

One strategy to improve our ability to simulate mixtures of complex fluids and collect information on an increasingly large range of length scales is to use a hierarchical approach, such as the multiscale modeling procedure that we present in this paper. 
In such a multiscaling scheme the system of interest is represented at various levels of coarse-graining and simulations are performed for each of these coarse-grained representations to enhance computational efficiency. In a second step, the information obtained from those simulations is collected and combined into a complete description, bridging information at all the length scales of interest.\cite{Plathe}

For any coarse-grained model to be capable of faithfully reproducing the desired characteristics of the system, the coarse level of description needs to include the essential features of the atomistic structure while averaging out the remaining degrees of freedom. Several coarse-graining procedures have been developed in the literature at various levels of ``coarseness". At the lowest level,  the united atom (UA) description, which represents groups of $\text{CH}_x$ units, with $x=1,2$, or $3$ into an effective UA along the chain, is a particularly useful representation,  for which potentials have been optimized to reproduce the structure for most polyolefins.\cite{UApapers} Higher levels of structure-based approaches simplify interactions, while remaining close to the chemical structure, facilitating scale hopping between the different levels\cite{Praprotnik}; however, these methods remain computationally demanding and do not adopt an extreme enough level of coarse-graining to mitigate the limitations of fully atomistic simulations in reaching the large length scales of phase separation.\cite{Baeurle} In order to simulate the increasingly large scale phenomena of interest to material scientists, an even coarser level of description is needed, while remaining formally related to the finer structure. 

The highest level of coarse-graining in our procedure, represents each polymer chain as a soft colloidal particle, centered at the polymer center-of-mass. Soft colloids interact through a Gaussian-like repulsive potential of the order of the polymer radius-of-gyration, $R_g$, which defines the characteristic length scale of this description. Large-scale behavior is easily generated in a ``mesoscale" simulation of the coarse-grained, mixture and information is collected on a scale equal to or larger than the polymer radius-of-gyration. Recently, we have shown that our mesoscale simulations capture the global structure of the polymer mixture providing information over much of the miscible region of the phase diagram.\cite{McCarty1}

Much attention has been given recently to relating mesoscopic\cite{Akkermans} or field-theoretical model parameters\cite{Sewell} to their more detailed atomistic representation. For example, Groot and Warren were able to express the repulsive soft potential parameters used in dissipative particle dynamics (DPD) in terms of the Flory-Huggins $\chi$ parameter, providing a bridge between DPD simulations and atomistic simulations, from which solubility parameters can be directly calculated.\cite{GROOT} This method has been successfully implemented in mesoscale simulations of biological membranes once the model is parameterized for specific molecular structures.\cite{GROOT2001} In our approach, the effective $\chi$ parameter is included directly in the model as an input parameter which determines the length scale of the of concentration fluctuations. This $\chi$ parameter provides a convenient link between an atomistic or united atom level description and a mesoscale picture, similar to other mesoscale modeling approaches;\cite{GROOT, Sewell, Titievsky} however, our approach has the advantage over other methods such as DPD in that the soft-core potential depends directly on the $\chi$ parameter and not on any phenomenological expression to bridge the two theories. In other words, once the $\chi$ parameter is specified, our multiscale modeling procedure will produce the expected phase behavior\cite{McCarty1}, and the task in practice is to determine this parameter for a given model.

In the current paper, we focus on bridging the length scales of binary homopolymer mixtures of different chemical architecture to provide a complete description of the static structure of the blend under a given set of thermodynamic conditions. 
Blends of polyolefins represent a good test case for our procedure since united atom parameters have been well optimized for these systems, and they provide a stringent test of the method, as subtle differences in chemical architecture, such as the extent of branching, greatly affects the global structure and miscibility. One advantage of our coarse-graining and multiscale modeling approaches is that since they are based on the formalism of the Ornstein-Zernike equation, the rescaling of the structure to recover the monomer level of description is straightforward. Furthermore, we have shown that the correct dynamics can be obtained directly from simulations of the coarse-grained structures by proper rescaling.\cite{Lyubimov} Thus, by performing simulations of mesoscopic particles and subsequently reinserting the relevant chemical details afterwards, we are able to obtain \emph{quantitative} information about large-scale properties.

Large scale information, obtained form mesoscale MD simulations, is later combined with local-scale united-atom simulations of the same system, using a multiscale formalism obtained from the solution of the Ornstein-Zernike equation, presented in this paper. In a spirit similar to the united atom simulations to which we test our approach, we initially take the high temperature limit as a reference state, and set $\chi=0$ in order to calculate monomer level total correlation functions. The model is then extended to incorporate thermal effects on the concentration fluctuation through the interaction parameter, $\chi$, which serves to bridge the gap between mesoscale and atomistic models.

This multiscale modeling approach, is an extension to polymer mixtures of our coarse-graining and multiscale modeling procedure for polymer melts.\cite{McCarty2} By comparison with united atom simulations of polymer mixtures, our approach shows good agreement between the two descriptions. Furthermore, our procedure has the advantage of being analytically solved, so that thermodynamic parameters defining the free energy potential appear explicitly in the formalism, making the whole procedure general and transferible to other systems in different thermodynamic conditions. The key advantage of our method lies in the fact that it provides a quantitative picture of how the local details such as the specific chemical architecture and thermodynamic conditions are connected to the observed, global properties. While atomistic and united atom simulations are limited in the maximum length scale that can be simulated, our approach easily provides information on a wide range of length scales,  once mesoscale simulations are combined with atomistic or monomer level simulations. The number of test systems presented in the paper is determined by the available united-atom simulations against which our procedure is tested; however, our procedure is directly applicable to other mixtures as well.

While the focus here is on the equilibrium structural properties of simple polyolefins, it is worth noting that many interesting systems could be investigated through our method and important problems approached. For example, the proposed multiscale procedure could be applied to study various aspects of binary blends, including critical behavior, mechanical properties and viscosity, the effects of non-random mixing, stiffness disparity, shape disparity, and different architectures, as well as the dynamics of demixing at equilibrium and under shear, interdiffusion in equilibrium and during phase separation, and  the different propensity for crystallization in the dynamics of miscible blends approaching the phase transition. As one of the polymers in the mixture crystalizes it is important to investigate, for example, the kinetics of interface formation and structure. Additionally the procedure could be extended to encompass mixtures including block and homo-copolymers, blends close to surfaces and in thin films, and mixtures of colloids and polymers. Although the dynamics of coarse-grained systems is unrealistically accelerated, we have recently proposed a new first-principle method to rescale the dynamics of the coarse-grained system to its atomistic value, enabling dynamical properties of these systems to be investigated with quantitative precision.\cite{Lyubimov} This method can be useful in evaluating long-time dynamics directly from the mesoscale simulation.

The paper begins with a brief overview of the theory for coarse-graining of polymer blends, from which we calculate the effective potential, $v(r)$, input to the mesocale molecular dynamic simulations of the coarse-grained system. Mesoscale simulations of the coarse-grained polymer mixture, where chains are represented as soft colloidal particles, are presented in the following section. Those simulations provide the large-scale information, which is combined with local-scale united-atom simulations by means of our multiscale modeling procedure. Next, we present our results in the form of the total correlation function, $h_{\alpha,\beta}(r)$ with $\alpha,\beta \in A,B$ for each pair of components, which is predicted and compared with data of full united-atom simulations. 
First, we present correlation functions of the mixture in the limit where \textbf{$\chi=0$} to demonstrate the tenor of our scale bridging approach. This is followed by an extension of the model to include finite temperature effects, where we present the change in the concentration structure factor upon inclusion of realistic values for the $\chi$ parameter. A brief discussion concludes the paper.

\section{Methodology}
\subsection{Effective Pair Potentials for Polymer Mixtures}
\label{SX:INTR}
\noindent An analytical coarse graining procedure for binary polymer blends was derived by Yatsenko, \emph{et al.}\cite{YAPRL}, to formally map the blend onto a mixture of soft-colloidal particles. Here, we briefly review the results of that work. The polymer species (A and B) in the blend are characterized by a radius of gyration, $R_{gA}$ and $R_{gB}$, a total density of monomers, $\rho$, and a segment length $\sigma_{A} = \sqrt{6/N_A}R_{gA}$ and $\sigma_{B} = \sqrt{6/N_B}R_{gB}$, where $N_A$ and $N_B$ are the number of chain units. The volume fraction of the A component is given by $\phi$ and the ratio of segment lengths is given by $\gamma=\sigma_A/\sigma_B$. The main physical quantity of interest in the study of complex liquids, is the pair distribution function, or the related total correlation function, as from this quantity any thermodynamic property of interest can be directly calculated.\cite{McQuarrie} Here the total correlation function has contributions from three terms: $h_{AA}(k)$, $h_{BB}(k)$, $h_{AB}(k)$, representing the three types on interaction of a mixture of particles A and B. Analytical expressions for the total correlation functions at the coarse grained level are obtained by solving a generalized Ornstein-Zernike (OZ) matrix relation

\begin{equation}
\textbf{H}(k)={\bf{\Omega} }(k)\textbf{C}(k) \left[ {\bf{ \Omega} }(k) +\textbf{H}(k) \right] \ ,
\label{EQ:BLND1}
\end{equation}
where \textbf{H}(k) is the matrix of total intermolecular partial correlation functions (pcfs), \textbf{C}(k) is the matrix of direct pcfs, and ${\bf \Omega} (k)$ represents the matrix of intramolecular pcfs.  Solving Equation \ref{EQ:BLND1} gives the relation for a generic pair\cite{YAJCP} $\{ \alpha, \beta \in (A, B) \} $
\begin{equation}
h^{cc}_{\alpha \beta}(k)=\left[\frac{\omega_{\alpha \alpha}^{cm}(k)\omega_{\beta \beta}^{cm}(k)}{\omega_{\alpha \alpha}^{mm}(k)\omega_{\beta \beta}^{mm}(k)}\right] h^{mm}_{\alpha \beta}(k),
\label{EQ:BLND2}
\end{equation}
where $\omega(k)$ for each type of chain is the intramolecular form factor, and the superscripts, $cc$, $cm$, or $mm$, denotes center of mass-center of mass, monomer-center of mass, or monomer-mononomer interactions, respectively.
The derivation of Equation \ref{EQ:BLND2} is an extension of the procedure outlined by Krakoviack, \emph{et al.}\cite{K2002} for homopolymers and formally connects the center of mass (cm) distribution functions to monomer-monomer (mm) distribution functions. Both the intramolecular form factors in Equation  \ref{EQ:BLND2}, namely the monomer-monomer $\omega^{mm}(k)$ and the monomer-cm $\omega^{cm}(k)$, are expressed in analytical forms.\cite{Yamakawa,DoiEd} Yatsenko, \emph{et al.} introduced a new analytical form  for the monomer total correlation functions, $h^{mm}_{\alpha \beta}$ by extending the PRISM-blend thread model of Tang and Schweizer\cite{TANG} to include asymmetries in local chemical structure and flexibility.\cite{YAPRL}

Following this procedure, analytical forms of $h^{cc}_{\alpha \beta}$, coarse-grained at the cm level, can be calculated from Equation \ref{EQ:BLND2}.  In real space they are given by:
\begin{eqnarray}
h_{AA}(r) & = & \frac{1-\phi}{\phi} I^{\phi}_{AA}(r) +\gamma^2I^{\rho}_{AA}(r) \nonumber \ , \\
h_{BB}(r) & = & \frac{\phi}{1-\phi} I^{\phi}_{BB}(r) +\frac{1}{\gamma^2} I^{\rho}_{BB}(r) \label{EQ:BLND3.1} \ , \\  
h_{AB}(r) & = & -I^{\phi}_{AB}(r) + I^{\phi}_{AB}(r) \nonumber \ , 
\end{eqnarray}
with
\begin{eqnarray}
& I^{\lambda}_{\alpha \beta}(r)=\frac{3}{4} \sqrt{ \frac{3}{\pi} } \frac{\xi'_{\rho}}{R_{g \alpha \beta}}\vartheta_{\alpha \beta1} \left(1-\frac{\xi^2_{c \alpha \beta}}{\xi^2_\lambda} \right)e^{-3r^2/(4R^2_{g \alpha \beta})} &\nonumber \\&-\frac{1}{2} \frac{\xi'_{\rho}}{r}\vartheta_{\alpha \beta2} \left(1-\frac{\xi^2_{c \alpha \beta}}{\xi^2_\lambda} \right)^2e^{R^2_{g \alpha \beta}/(3 \xi^2_\lambda)}&  \\
& \times \left[e^{r/\xi_\lambda}\mbox{erfc}\left(\frac{R_{g \alpha \beta}}{\xi_\lambda \sqrt{3}}+\frac{r\sqrt{3}}{2R_{g \alpha \beta}}\right)-e^{r/\xi_\lambda}\mbox{erfc}\left(\frac{R_{g \alpha \beta}}{\xi_\lambda \sqrt{3}}-\frac{r\sqrt{3}}{2R_{g \alpha \beta}}\right)\right] \nonumber \, 
\label{EQ:BLND4}
\end{eqnarray} 
and 
\begin{eqnarray}
\vartheta_{\alpha \beta1} &=& \frac{\left(1-\frac{\xi^2_{c \alpha \alpha} \xi^2_{c \beta \beta} }{\xi^2_{c \alpha \beta}\xi^2_\lambda}\right)}{\left(1-\frac{\xi^2_{c \alpha \beta}}{\xi^2_\lambda}\right)} \ , \\
\vartheta_{\alpha \beta2} &=&\frac{\left(1-\frac{\xi^2_{c \alpha \alpha}}{\xi^2_\lambda}\right)\left(1-\frac{\xi^2_{c \beta \beta}}{\xi^2_\lambda}\right)}{\left(1-\frac{\xi^2_{c \alpha \beta}}{\xi^2_\lambda}\right)^2} \ ,
\label{EQ:BLND5}
\end{eqnarray}
where $\xi_{c \alpha \beta}$ is the average correlation hole length scale, and $R_{g \alpha \beta} \equiv \sqrt{(R^2_{g \alpha} +R^2_{g \beta} )/2}=\xi_{c \alpha \beta}\sqrt{2}$ is an average radius-of-gyration. Here, $\xi_\lambda \in \{\xi_\rho,\xi_\phi\}$ identifies the length scale for the density and concentration fluctuation correlations. The concentration fluctuation of a polymer blend is defined as
\begin{equation}
\xi_\phi = \frac{\sigma_{AB}}{\sqrt{24 \chi_s (1-\frac{\chi}{\chi_s}) \phi (1-\phi)}} \ , 
\label{EQ:BLND6}
\end{equation}
where we introduce the miscibility parameter $\chi$ for which the quantity $\chi/\rho$ is analogous to the Flory-Huggins interaction parameter\cite{YAJCP} such that $\chi \propto 1/T$. The concentration fluctuation, $\xi_\phi$, diverges at the spinodal temperature where $\chi =\chi_s$. The average segment length is $\sigma^2_{AB} = \phi \sigma^2_B +(1-\phi)\sigma^2_A$, while the density fluctuation, $\xi_\rho$, is defined as $\xi^{-1}_{\rho \alpha \beta}=\pi \rho \sigma^{2}_{\alpha \beta}/3 + \xi^{-1}_{c \alpha \beta}$ and $\xi^{\prime}_\rho=3/(\pi \rho \sigma^2_{AB})$. Although for compressible blends, there is, in principle, more than one $\chi$ parameter resulting from the distinct partial osmotic compressibilities of a two component blend\cite{Schweizer1995}, in this work we adopt a single interaction parameter which determines the length scale of concentration fluctuations. The choice of a single adjustable parameter is appropriate in this case, since the aim is to obtain quantitative agreement with experiment, and the adoption of a single ``effective" $\chi$ maintains a straightforward connection to SANS experiments. The notion of extracting a single effective $\chi$ parameter was also invoked Dudowicz \emph{et al.} in a general lattice cluster theory analysis of compressible blends,\cite{Dudowicz} and by Schweizer in connecting the PRISM blend approach with methods used in experimental SANS analysis.\cite{SchweizerMacro}  While we initially set $\chi=0$, the extension to more realistic models is achieved by using  the experimental $\chi$ parameter within this general framework, which has been shown to represent well both upper and lower critical phase diagrams.\cite{YAJCP} Results presented in Section III B of this paper are examples of both kinds of systems. In the athermal regime, however, the theory is presently being implemented to reproduce quantitatively and in a self-consistent way, the experimental values of $\chi_{ath}$ \cite{workinprog}.

Finally, it should be noted that here we are only interested in capturing the global structure and thus the use of the simple analytic PRISM thread result is adequate as it accurately captures the structure of the polymer at distances greater the $R_g$ and correctly predicts the correlation hole. While the PRISM thread result does not capture local effects such as solvation shells, such detail is not relevant at the coarse-grained level since at this level, local interactions are averaged out. Local packing in the structural g(r) is introduced later using the multiscale modeling procedure discussed later, where local scale information from UA MD simulations is paired with global information obtained from course-grained simulations.

By adopting the hypernetted-chain (HNC) closure relation, the effective pair interaction potential, $v^{cc}_{\alpha \beta}(r)$, is connected to the pcfs by the relationship
\begin{equation}
(k_B T)^{-1} v^{cc}_{\alpha \beta}(r)=h^{cc}_{\alpha \beta}(r)-\ln [1+h^{cc}_{\alpha \beta}(r)] - c^{cc}_{\alpha \beta}(r) \ ,
\label{EQ:BLND7}
\end{equation}
where $c^{cc}_{\alpha \beta}$ is the direct pcf for a mixture of soft colloidal particles, defined in reciprocal space as
\begin{eqnarray}
c^{cc}_{\alpha \alpha}(k) &=& \frac{1}{\rho_{c,\alpha}}-\frac{S^{cc}_{\beta \beta}(k)}{(\rho_{c,\alpha} +\rho_{c, \beta})|\bold{S}_{cc}(k)|} \ , \nonumber \\
c^{cc}_{\alpha \beta}(k) &=&\frac{S^{cc}_{\alpha \beta}(k)}{(\rho_{c,\alpha} +\rho_{c, \beta})|\bold{S}_{cc}(k)|} \ . 
\label{EQ:BLND8}
\end{eqnarray}
The chain density, $\rho_{c,\alpha}$, of chain type, A, is given as $\rho_{c,A}=\phi \rho/N_A$, and for chain type, B, as $\rho_{c,B}=(1-\phi) \rho/N_B$.
For a binary mixture the static structure factors, $S^{cc}_{\alpha \beta}$,  are given by 

\begin{eqnarray}
S^{cc}_{AA}(k) &=& \phi +\phi^2 \rho_{ch} h_{AA}(k) \nonumber \ , \\
S^{cc}_{BB}(k) &=& 1- \phi +(1- \phi )^2 \rho_{ch} h_{BB}(k) \nonumber \ , \\
S^{cc}_{AB}(k) &=& \phi (1-\phi ) \rho_{ch} h_{AB}(k) \nonumber \ , \\
\label{EQ:BLND9}
\end{eqnarray}
where the total chain density $\rho_{ch}=\rho/N$, and $|\bold{S}_{cc}(k)|=S^{cc}_{AA}(k)S^{cc}_{BB}(k)-[S^{cc}_{AB}(k)]^2$ is the determinant of the mesoscopic static structure factor matrix. 
Whereas at the monomer level of description, molecular closures are required to ensure the correct scaling of the $\chi$ parameter\cite{SchweizerYethiraj}, the use of the site level HNC is appropriate here because at this level of coarse-graining, the polymers are treated as a simple liquid of soft-colloids.\cite{LouisPRL}

By inserting Equations \ref{EQ:BLND3.1} and \ref{EQ:BLND8} into equation \ref{EQ:BLND7}, the effective pair potentials are calculated, which are input to the mesocale simulations of  the coarse grained binary mixture. The potential so obtained explicitly depends on the structural parameters of the polymer, such as $\rho$, $N$, $\phi$, and $R_g$. Since these parameters enter into the UA description, they do not represent additional parameters needed in moving to a higher level of coarse-graining. In other words, the potential is state dependent, being a free energy obtained from the monomer frame of reference;  however, it is fully transferable to different systems since it is analytically derived in a well-defined manner. In addition to these structural parameters, there is the additional parameter, $\chi$, that enters into the mesoscale description through Equation \ref{EQ:BLND6}, and describes the monomer-specific interactions that drive phase separation. Once these parameters are defined, mesoscale simulations may be performed and structural properties calculated for any system of interest.

\subsection{Mesoscale Simulations of Coarse-Grained Polymer Mixtures}
In continuing our multi-scaling approach, mesoscale simulations (MS) of various binary polymer mixtures with $\chi=0$ were performed using the effective potential calculated in the previous section. These simulations provide a reference system representative of the case in which the blend is well mixed and fluctuations are small. Systems investigated were blends of polyethylene (PE), polyisobutylene (PIB), and polypropylenes in their head-to-head (hhPP), isotactic (iPP), and syndiotactic (sPP) forms. Table \ref{TB:1} shows the simulation parameters used in both UA and MS descriptions for each polyolefin blend studied.

\begin{table}[h!b!p!] 
\centering
\caption{Simulation Parameters for Polyolefin Blends ($N_A = N_B = 96)$}
\begin{tabular}{cccccc}
  \hline \hline
Blend [A/B] & $\phi$ & $\rho$ [sites/$\AA$] & $R_{gA}$ [$\AA$] & $R_{gB}$ [$\AA$]  & $\gamma$ \\
\hline 
hhPP/sPP & 0.50 & 0.0332 & 12.18 & 13.87 & 1.14 \\
hhPP/PE & 0.50 & 0.0332 & 12.32 & 16.48 & 1.34 \\
PIB/PE & 0.50 & 0.0343 & 9.76 & 16.38 & 1.68 \\
PIB/sPP & 0.50 & 0.0343 & 9.76 & 13.78 & 1.41 \\
iPP/PE & 0.75 & 0.0328 & 11.33 & 16.69 & 1.48 \\
hhPP/PIB & 0.50 & 0.0343 & 12.41 & 9.69 & 1.28 \\
\hline \hline
\label{TB:1}
\end{tabular}
\end{table}

Our MS MD simulations were performed on systems of point particles evolving in the microcanonical ($N$, $V$, $E$), ensemble. Initially all particles were placed on a cubic lattice with periodic boundary conditions, where the type of particle ($A$/$B$) occupying a particular lattice site was determined at random. The potential and its corresponding derivative were entered as tabulated inputs to the program, and linear interpolation was used to determine function values not supplied as the algorithm proceeds. Each site was given an initial velocity pooled from a Mersenne Twister random number generator,\cite{Matsumoto} and the system subsequently was evolved using a velocity Verlet integrator. We used reduced units such that all the units of length were scaled by $R_{gAB}$ ($r^*=r/R_{gAB}$) and energies were scaled by $k_BT$.

Equilibrium was induced by a quenching procedure, in which velocities were rescaled at regular intervals to maintain the desired temperature. Proper equilibration was verified by observing a Maxwell-Boltzmann distribution of velocities, a steady temperature, a stabilized Boltzmann H-theorem function, and a decayed translational order parameter. Once velocity rescaling was discontinued during the production stage, we monitored the fluctuating steady temperature to ensure the system remains in equilibrium throughout the duration of the simulation. 

During the production stage, trajectories were collected over a traversal of $\sim 8 R_g$. A typical MS MD simulation included $\sim 4000$ particles, evolving for 30,000 computational steps. Several simulations were run using the LONI TeraGrid system\cite{TeraGrid} to facilitate performing numerous simulations at a time. As a benchmark, a typical run on a single CPU workstation took $\sim 5$ h of wall clock time; however, since our codes are not yet subject to any parallelization process, we stress that this represents an underestimate of the potential gain in computational time as opposed to running full atomistic or UA simulations. 

\begin{table}[h!b!p!] 
\centering
\caption{Mesoscale Simulation (MS-MD) Particle Number and Box Dimension Compared to UA Box Dimension. All UA simulations are for 1600 chains.}
\begin{tabular}{cccc}
  \hline \hline
Blend [A/B] & Particles & $L_{MS}$ [$\AA$]  & $L_{UA}$ [$\AA$] \\ 
\hline 
hhPP/sPP & 2744 & 199.07 & 166.61  \\
hhPP/PE & 5324 & 246.21 & 166.61 \\
PIB/PE & 4096 & 218.66 & 164.91 \\
PIB/sPP & 5488 & 245.19 & 164.91 \\
iPP/PE & 1728 & 168.57  & 167.27 \\
hhPP/PIB & 3456 & 230.73 & 164.91 \\
\hline \hline
\label{TB:2}
\end{tabular}
\end{table}

Mesoscale simulations provide the center-of-mass total correlation functions that describe the polymer mixtures on the large scale and are readily calculated from the simulation coordinates. As an exmple we show in Figure 1(a) the plot of $h^{cc}_{AA}(k)$ for a 50:50 mixture of hhPP/sPP ($\chi =0$). Data from mesoscale simulations and theoretical predictions are compared against united atom simulations for the center-of-mass total correlation functions and show an excellent agreement. Analytical theory, mesoscale simulations, and united atom simulations are all consistent in depicting the structure of the fluid on the length scale of the polymer radius of gyration and larger. Although all of the structural information is already contained in the analytic expression, the MS-MD simulations are useful as they can provide further information, for example, on the dynamics of the system close or far from equilibrium, both of which could be in principle investigated with our coarse-grained systems.

%\begin{figure}[h!]
%\centering
%\includegraphics[scale=.6]{Figure1_color.pdf}
%\caption{(a) Plot of $h^{cc}_{AA}(k)$ for hhPP/sPP obtained from mesoscale simulation (open red circles). Comparison with UA data (filled circles) and theoretical predictions (solid line) shows quantitative agreement. (b) Plot of $h^{mm}_{AA}(k)$ calculated using the inverse mapping procedure, Equation \ref{EQ:BLND10}, (open red circles) compared to data from the full UA MD simulation (solid circle).}
%\label{FG:MS}
%\end{figure}
\begin{figure}[h!]
\centering
\includegraphics[scale=.55]{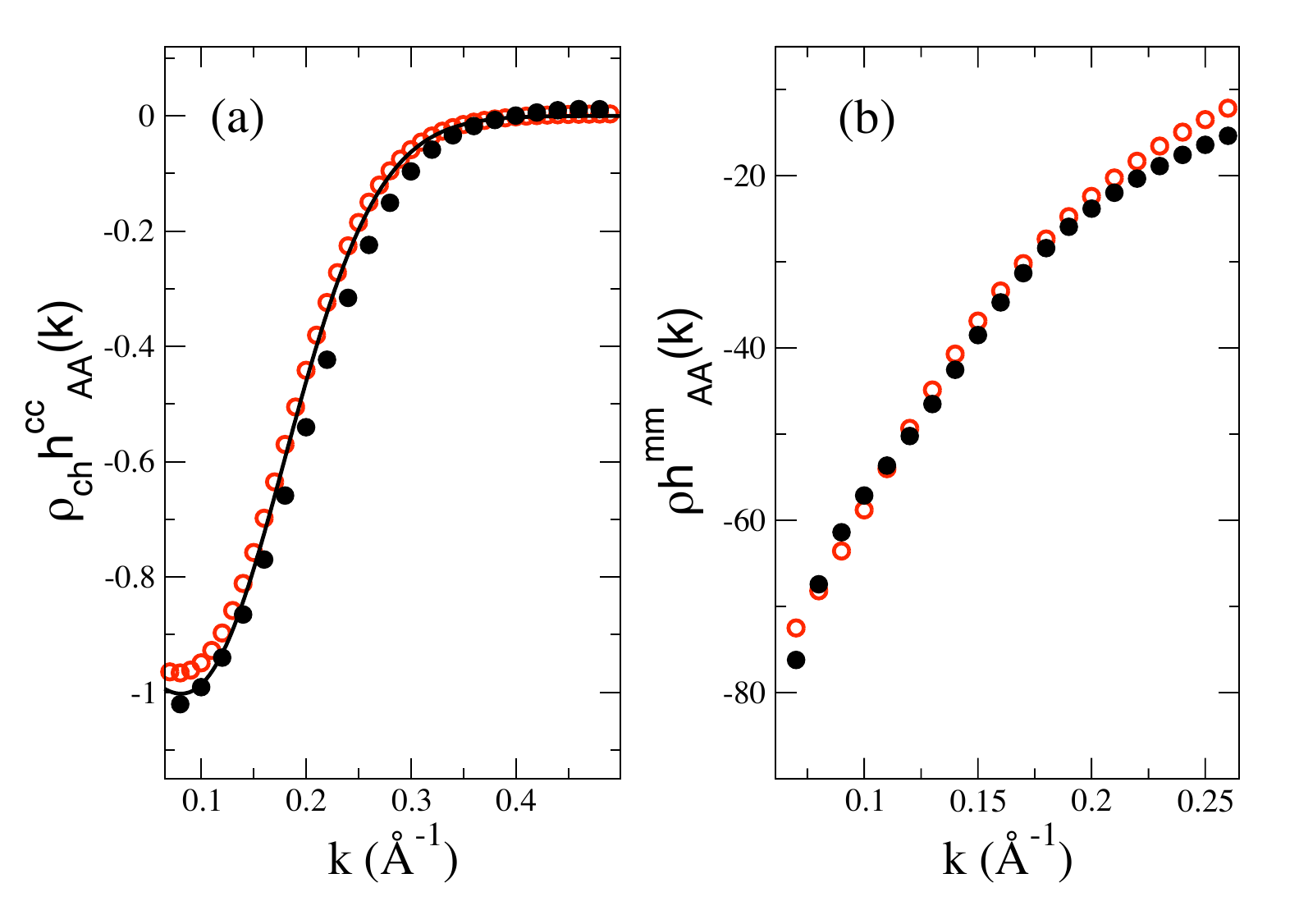}
\caption{(a) Plot of $h^{cc}_{AA}(k)$ for hhPP/sPP obtained from mesoscale simulation (open red circles). Comparison with UA data (filled circles) and theoretical predictions (solid line) shows quantitative agreement. (b) Plot of $h^{mm}_{AA}(k)$ calculated using the inverse mapping procedure, Equation \ref{EQ:BLND10}, (open red circles) compared to data from the full UA MD simulation (solid circle).}
\label{FG:MS}
\end{figure}

However, while $h^{cc}(k)$ describes the center of mass correlations between particles, when we are concerned with the structure of the liquid on the local scale, it is necessary to look at the monomer total correlation functions. Although the large scale information is completely provided by the mesoscale information, the information on the local-scale structure is 
averaged out during the coarse-graining. To take advantage of the gain in computational time given by mesoscale simulation, it is convenient to combine the large scale information from the latter with a short united atom simulation that describes the local structure. The two are merged together through a multiscale modeling procedure as presented in the next section. The advantage in performing this two-steps procedure is that the simulation with atomistic resolution can now be limited to short simulation runs and to a small ensemble of molecules, as the needed large-scale, long-time information comes from the fast mesoscale simulations.

\subsection{Recovering the Monomer Level Description: Implementation of the Multiscale Modeling Procedure for Polymer Blends}
\label{SX:MSBLND}

In our multiscale modeling we combine the mesoscale simulation, which correctly captures the large scale, global structure, with UA simulations, needed to determine the local properties of the polymer. 
By this procedure we aim at bridging the different length scales characteristic of the mixture and obtaining a complete description of the polymer blend structure at a defined temperature and composition. 

The key to this procedure lies in extracting the monomer-monomer correlation function, $h^{mm}(k)$ from the coarse-grained correlation function, $h^{cc}(k)$, by inverse mapping, using the reciprocal relation:
\begin{equation}
h^{mm}_{\alpha \beta}(k)=\left[\frac{\omega_{\alpha \alpha}^{mm}(k)\omega_{\beta \beta}^{mm}(k)}{\omega_{\alpha \alpha}^{cm}(k)\omega_{\beta \beta}^{cm}(k)}\right]h^{cc}_{\alpha \beta}(k) \ .
\label{EQ:BLND10}
\end{equation}
\noindent Here, $h^{cc}(k)$ is obtained from the mesoscale simulation, while the intramolecular form factors, $\omega(k)$, entering Equation \ref{EQ:BLND10} are calculated directly from short UA MD simulations. Since UA-MD simulations are needed only to obtain the intramolecular distributions, these simulations need only be for system box sizes on the order of the radius of gyration, which is a significant advantage over obtaining the global structure from UA-MD alone which requires much larger box sizes to accurately calculate the intermolecular distributions. 
The calculated $h^{mm}(k)$ is valid only for small k values, where the coarse-grained description applies, and begins to diverge as $\omega^{cm}(k)$ approaches zero at large k. This is shown in Figure 1(b), where we compare directly $h^{mm}(k)$ calculated using Equation \ref{EQ:BLND10} with UA-MD data of the same system.

To include local scale information, the total correlation function $h^{mm}(k)$ at small $k$, obtained following the described procedure, is combined with $h^{mm}(k)$ from UA simulation for large values of $k$. The key point in the procedure is to define a cut-off length, $k_{cut}\propto 1/r_{cut}$, at which the two descriptions are to be combined. 

Because the local scale information is averaged out in the coarse-grained description, it is important to ensure that the whole intramolecular description is accounted for by the UA simulation. To estimate the length scale at which local intramolecular contributions become negligible, we evaluate the fraction of \emph{intra} to \emph{total} site/site contacts. For a given component of type $\alpha$ in the mixture, this function is defined as

\begin{equation}
f_{s\alpha \alpha}(r)=\frac{N_{s\alpha \alpha}(r)}{N_{total}(r)} \ ,
\label{EQ:fs}\
\end{equation}

\noindent where $N_{s \alpha \alpha}(r)$ is the number of $\alpha$ type intramolecular contact sites
\begin{equation}
N_{s \alpha \alpha}(r)=4 \pi \rho \int^r_0 (r')^2\omega_{\alpha \alpha}^{mm}(r')\,dr' \ ,
\label{EQ:Ns}\
\end{equation}
while the total number of site/site contacts is given by 
\begin{equation}
N_{total}(r)=\frac{4}{3} \pi \rho r^3.
\label{EQ:Nt}
\end{equation}
In our procedure, the integral in Equation \ref{EQ:Ns} is computed numerically using UA data for $\omega_{\alpha \alpha}^{mm}(r)$. 

Selecting the correct cut-off distance is an important step of the procedure. Choosing a larger cutoff radius $r_{cut}$, smaller $k_{cut}$, includes more information from UA simulations, increasing the precision of the calculation, but it increases the computational time, partially defeating the purpose of adopting a multiscale procedure. In the case of polymer melts we observed that a value of $f_s(r)=0.025$ gives good precision in the calculated $h^{mm}(r)$ when compared with full united atom simulation data, while retaining a reasonable speeding up of the calculations.\cite{McCarty2} For polymer blends we adopt the 
same criterium and we evaluate \textit{a posteriori} the agreement obtained for the blend samples analyzed in this paper. In Table \ref{TB:NtBlnd} we report, for both AA and BB type interactions, the effective radius, $r_{cut}$, for combining simulations, which lies in the intermediate length scale on the order of a few $R_g$ units and is unique for each blend. Also reported is the total number of site/site contacts within the radius determined by the value $f_s(r)=0.025$, evaluated using Equation \ref{EQ:Nt} at $r=r_{cut}$. Dividing $N_{total}$ by the number of united atoms per chain (96 in our case) gives the number of chains, $n$, needed in a UA simulation to produce the statistical information necessary to obtain the local structure. This is important since it determines how ``short" the UA-MD simulations must be without losing pertinent information about local correlations.

Large-scale and local-scale information were combined at the chosen $k_{cut}= 2\pi/r_{cut}$, and the procedure was repeated for each of the mixtures shown in Table \ref{TB:1}. The correlation coefficient between $h^{mm}(r)$ determined from our multiscaling procedure and from full UA simulations is calculated for the case when the multiscale simulations are combined at $f_s=0.025$. These values are also presented in Table \ref{TB:NtBlnd}, showing that once $r_{cut}$ is defined, the multiscale procedure provides an accurate way of obtaining the total correlation function for binary blends while circumventing the need for prohibitively large atomistic simulations. For the case of AB type interactions, an average radius between AA and BB data is used to determine where the simulations should be combined. While this approach allows one to explore a large range in the degree of polymerization, the method becomes impractical for liquids of long chains, for which intramolecular effects will remain long-ranged, since the cutoff length scale occurs on the order of $R_g$. Moreover, for long chains, entanglement effects enter the dynamics and have to be accounted for. Therefore, for large N systems it is advisable to include a third, intermediate, level of coarse-graining, with a cut-off of the order of (and larger than) the persistence length and the entanglement length scale.

\section{Results}
\subsection{Total Pair Correlation Functions of Athermal Reference Blends}
The method just discussed gives a systematic way of merging simulations to optimize the tradeoff between the gain in accuracy due to inclusion of UA simulation data and the gain in efficiency due to the coarse grained mesoscopic picture.  
This procedure works well as it yields total correlation functions in excellent agreement with UA simulations at a reduced computational cost. 

\begin{table}[h!b!p!] 
\centering
\caption{Effective Radius for Combining Simulations and the Number of Total Site/Site Contacts Evaluated at $f_s(r) = 0.025$}
\begin{tabular}{cccccc}
  \hline \hline
System & Type & $r_{cut}[\AA]_{f_s=0.025}$ & $N_{total}$ & $n$ & Corr. Coeff \\
\hline
hhPP/sPP & AA & 29.1 & 3427 & 36 & 0.9999 \\
hhPP/sPP & BB & 28.7 & 3288 & 34 & 0.9999 \\
hhPP/PE96 & AA & 28.7& 3288 & 34 & 0.9998 \\ 
hhPP/PE96 & BB & 27.6 & 2924 & 31 & 0.9999 \\
PIB/PE96 & AA & 28.8 & 3432 & 36 &  0.9999  \\
PIB/PE96 & BB & 27.2 & 2891 & 30 &  0.9999 \\
PIB/sPP & AA & 29.1 & 3536 & 37 & 0.9995 \\
PIB/sPP & BB & 29.6 & 3721 & 39 & 0.9999 \\
iPP/PE96 & AA & 29.3 & 3458 & 36 & 0.9999 \\
iPP/PE96 & BB & 27.6 & 2891 & 30 & 0.9998  \\
\hline \hline
\label{TB:NtBlnd}
\end{tabular}
\end{table}

\begin{figure}[h!]
\centering
\includegraphics[scale=.55]{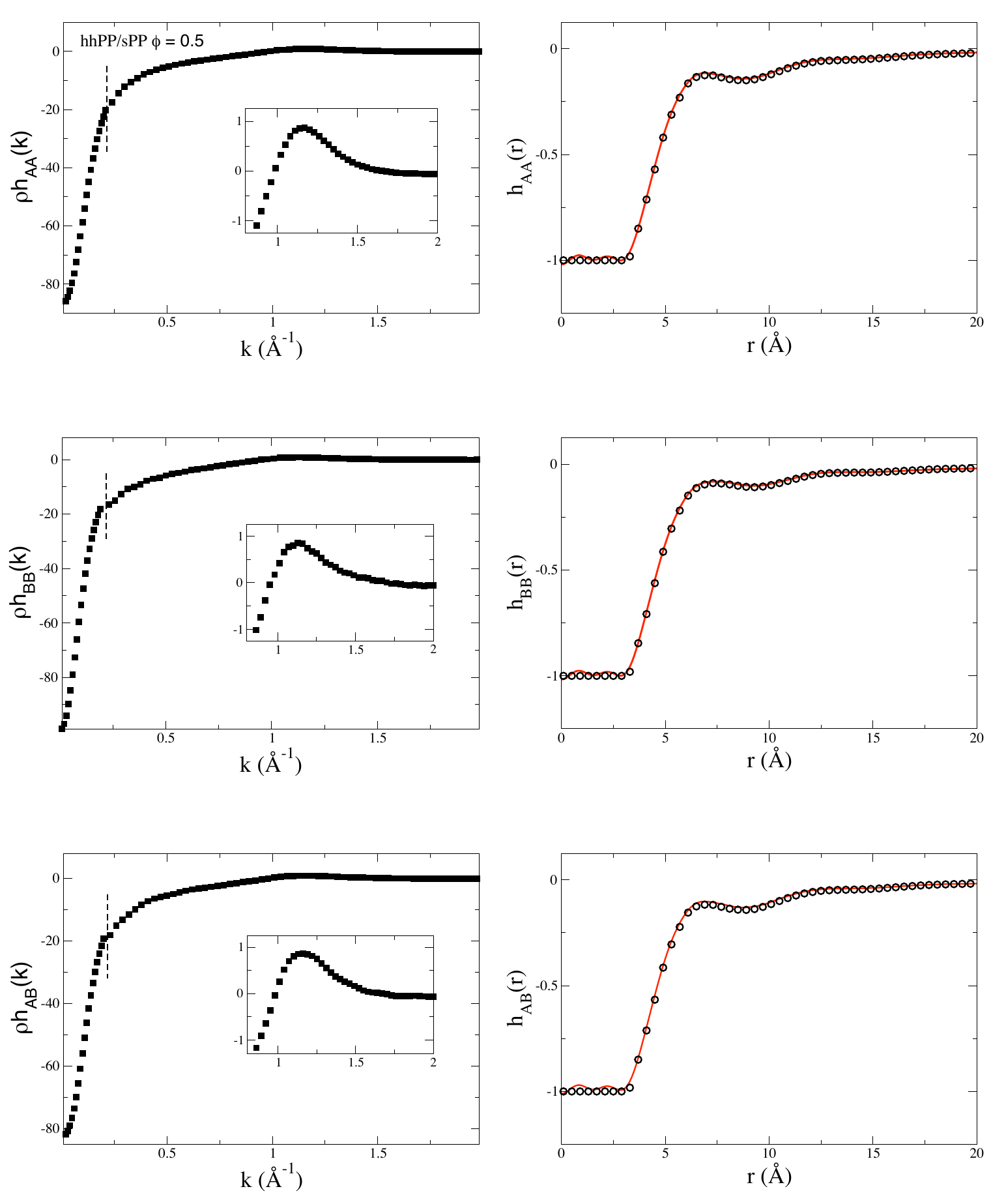}
\caption{\small{(Left) Multiscale modeling: The left panels show the total correlation function, $h^{mm}(k)$, for AA (top), BB (middle), and AB (bottom) interactions, for a mixture of 50:50 hhPP/sPP. The data over the range of small k values was obtained by mesoscale simulation, whereas over the large k range it was obtained by UA MD simulation. The inset depicts the local structure. The dashed line indicates the value at which the two simulations were combined. (Right) The correlation function, $h^{mm}(r)$, after Fourier transform (solid red line) is compared with results from the full UA MD simulation (open symbols).}}
\label{FG:HHPP_sPP}
\end{figure}

\begin{figure}[h!]
\centering
\includegraphics[scale=.55]{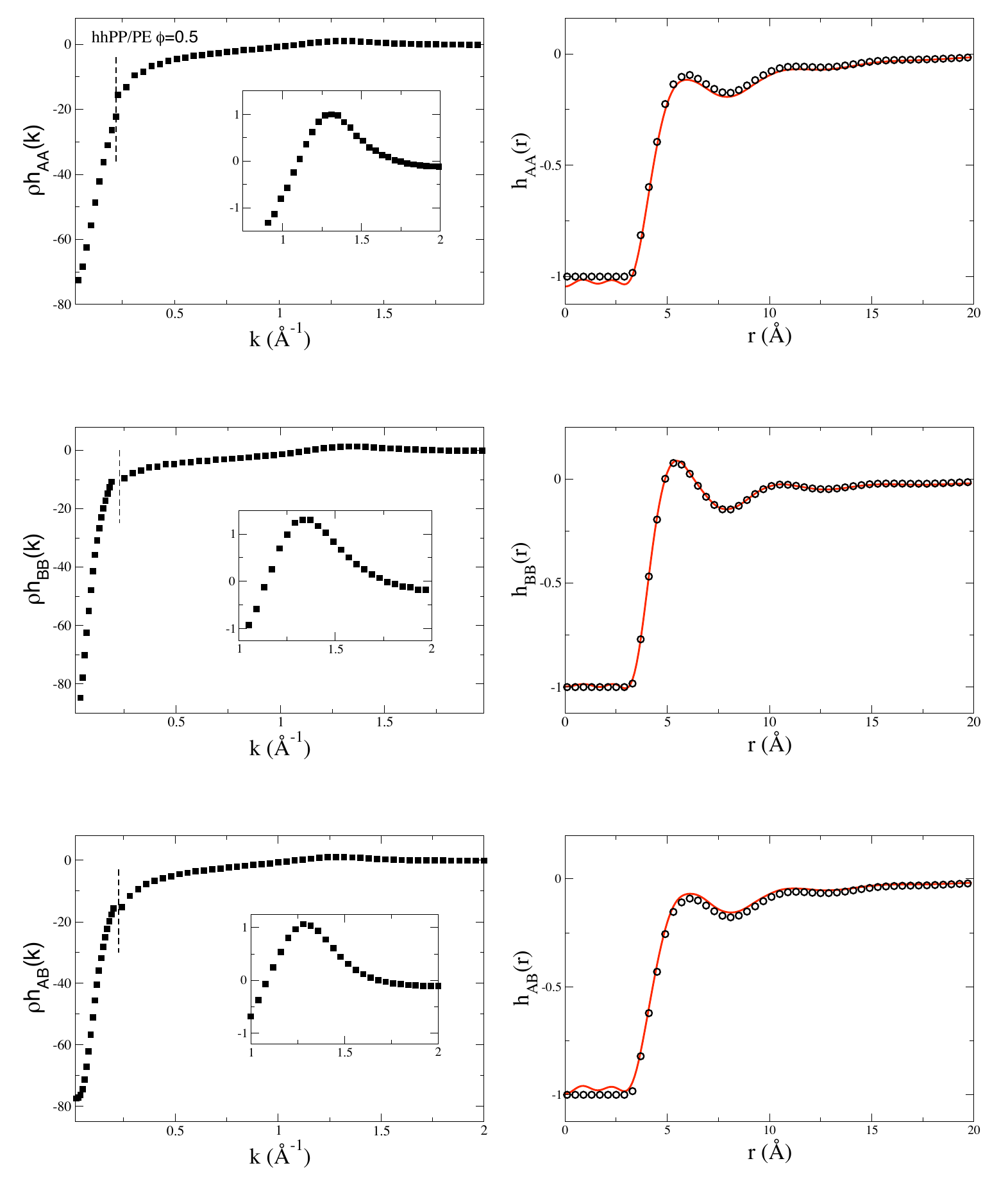}
\caption{Same as in Figure \ref{FG:HHPP_sPP} for a blend of 50:50 hhPP/PE.}
\label{FG:HHPP_PE}
\end{figure}

\begin{figure}[h!]
\centering
\includegraphics[scale=.55]{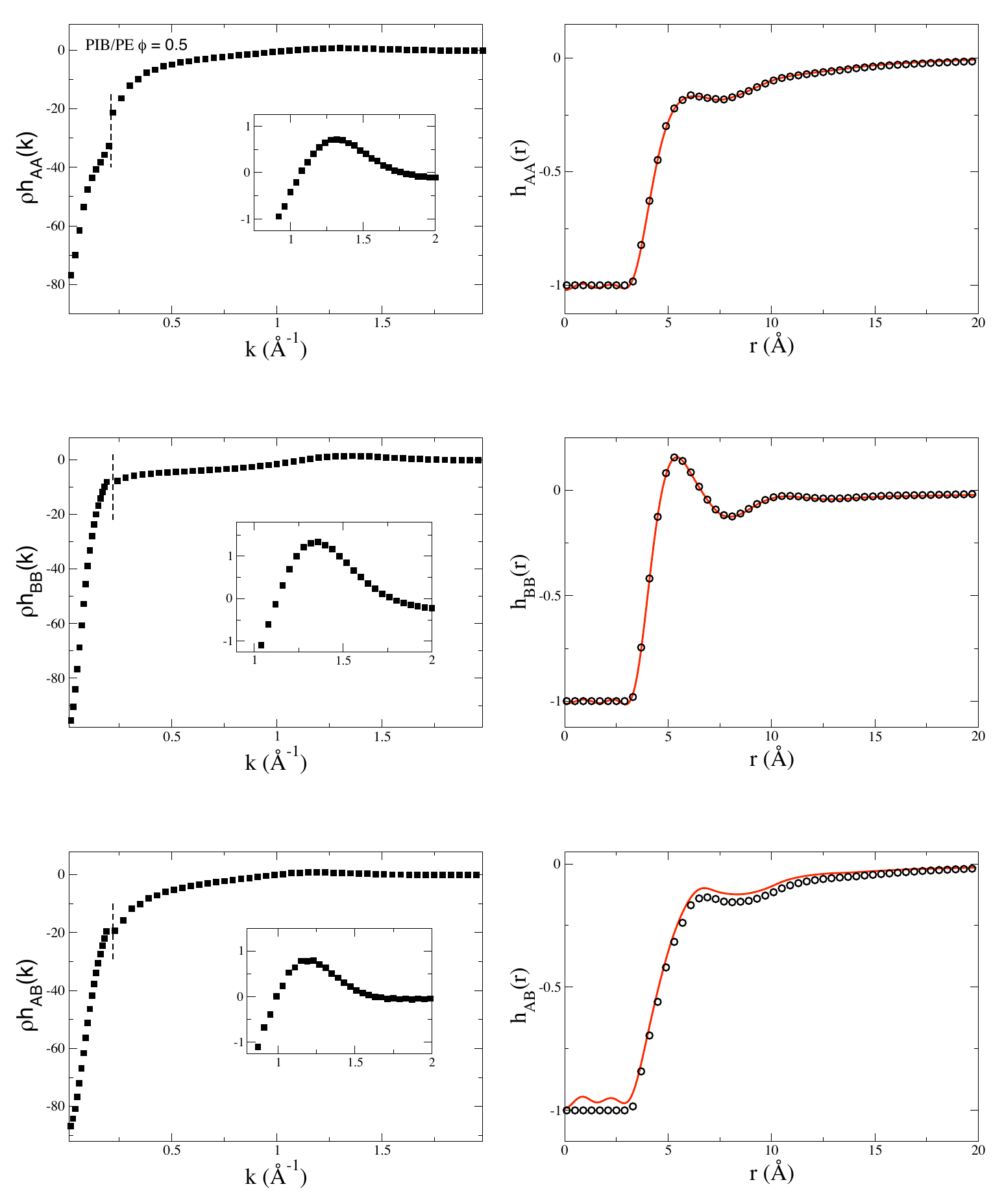}
\caption{Same as in Figure \ref{FG:HHPP_sPP} for a blend of 50:50 PIB/PE.}
\label{FG:PIB_PE}
\end{figure}

\begin{figure}[h!]
\centering
\includegraphics[scale=.55]{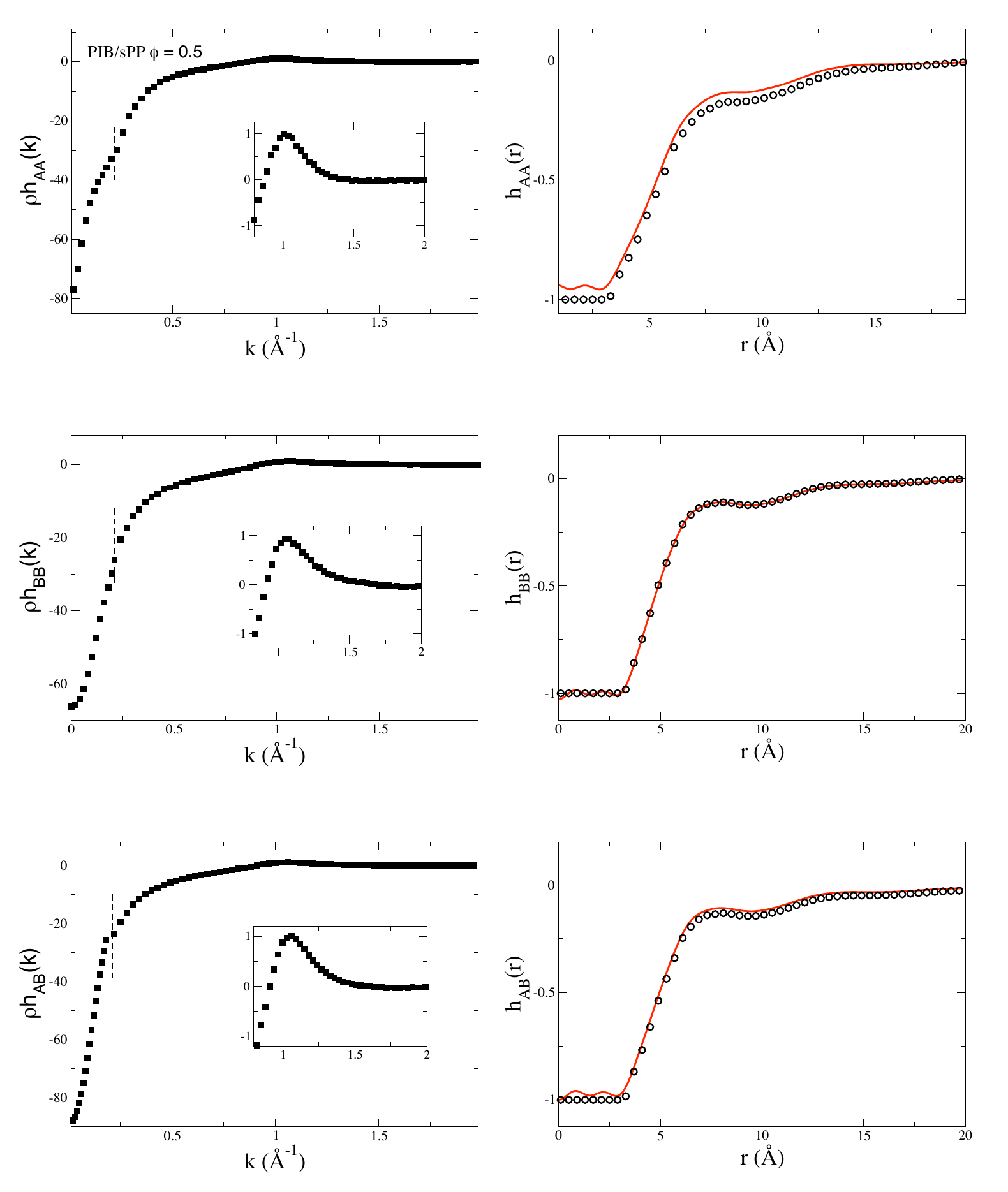}
\caption{Same as in Figure \ref{FG:HHPP_sPP} for a blend of 50:50 PIB/sPP.}
\label{FG:PIB_sPP}
\end{figure}

\begin{figure}[h!]
\centering
\includegraphics[scale=.55]{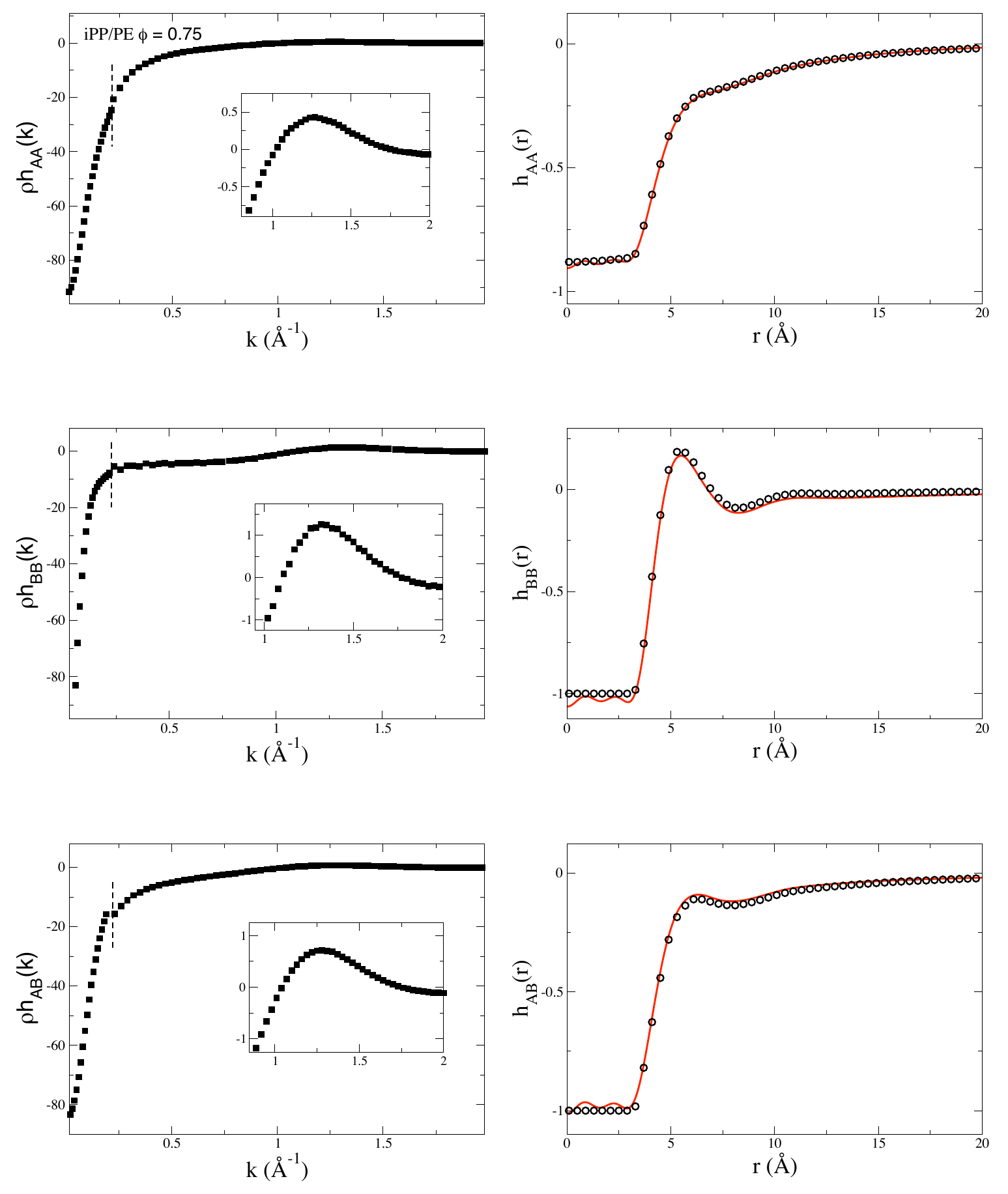}
\caption{Same as in Figure \ref{FG:HHPP_sPP} for a blend of 75:25 iPP/PE.}
\label{FG:iPP_PE}
\end{figure}

Results from the multiscale procedure are reported in Figures \ref{FG:HHPP_sPP} - \ref{FG:iPP_PE}.
In Figure  \ref{FG:HHPP_sPP}, the left panels show the total correlation function, $h^{mm}(k)$, obtained by combining UA simulations for the short length scales (large k) with mesoscale simulations for the long length scales (small k), for the first mixture in Table \ref{TB:1}, hhPP/sPP. The vertical dashed line in the left panels represents the cut-off length scale at which the two simulations are combined, which corresponds to $f_s=0.025$.
The insert in the left panels depicts the peak representing the local structure, which is accurately determined by UA simulation. This peak depends on the geometry of the monomeric structure and changes as a function of the type of polymers involved in the mixture. 

The right panels in Figure \ref{FG:HHPP_sPP} show $h^{mm}(r)$, i.e. the Fourier transform of the corresponding total correlation function, $h^{mm}(k)$.  As in the case for melts, we adopted a sampling step of $\Delta k =0.01$ in reciprocal space, as this is of the same order as the discontinuity in $h^{mm}(k)$, which results from joining the two simulations. Because of the amplitude of the chosen sampling step, there is no effect on the Fourier transformed function $h^{mm}(r)$. The pair distribution function calculated from the multiscale modeling procedure is identical to the coarse-grained analytical expression in the large-scale regime, but also includes the local scale solvation shells, which come from UA-MD and are not captured by the PRISM thread expressions. Both pieces of information are needed to provide the complete structural and thermodynamic picture of the system. 

Figure \ref{FG:HHPP_sPP} illustrates the spirit of our multi-scale modeling approach in which independent simulations representing the same system at different levels of coarse-graining can be combined to provide a complete description of the polymer. 
Analogous plots for the hhPP/PE and the PIB/PE mixture are reported in Figure \ref{FG:HHPP_PE} and Figure \ref{FG:PIB_PE}, respectively. Finally, results for the two mixtures of PIB/sPP and iPP/PE are presented in Figures \ref{FG:PIB_sPP} and \ref{FG:iPP_PE}.
 All systems in real space show quantitative agreement with UA-MD data but are obtained at a much more efficient computational time than running the full UA MD simulation. Furthermore, the procedure to obtain the pair correlation functions is entirely analytical, hence we do not utilize any optimization procedure or numerical fitting scheme to obtain consistency between the two descriptions, and thus the method is fully transferable. Figures \ref{FG:HHPP_sPP}-\ref{FG:iPP_PE} demonstrate the versatility of the approach for studying mixtures of polymers with subtly different chemical architecture, since the same multiscale procedure is applied in all cases. 
 
%\begin{figure}[h!]
%\centering
%\includegraphics[scale=1.0]{Figure2_color.pdf}
%\caption{\small{(Left) Multiscale modeling: The left panels show the total correlation function, $h^{mm}(k)$, for AA (top), BB (middle), and AB (bottom) interactions, for a mixture of 50:50 hhPP/sPP. The data over the range of small k values was obtained by mesoscale simulation, whereas over the large k range it was obtained by UA MD simulation. The inset depicts the local structure. The dashed line indicates the value at which the two simulations were combined. (Right) The correlation function, $h^{mm}(r)$, after Fourier transform (solid red line) is compared with results from the full UA MD simulation (open symbols).}}
%\label{FG:HHPP_sPP}
%\end{figure}

%\begin{figure}[h!]
%\centering
%\includegraphics[scale=1.0]{Figure3_color.pdf}
%\caption{Same as in Figure \ref{FG:HHPP_sPP} for a blend of 50:50 hhPP/PE.}
%\label{FG:HHPP_PE}
%\end{figure}

%\begin{figure}[h!]
%\centering
%\includegraphics[scale=1.0]{Figure4_color.pdf}
%\caption{Same as in Figure \ref{FG:HHPP_sPP} for a blend of 50:50 PIB/PE.}
%\label{FG:PIB_PE}
%\end{figure}

%\begin{figure}[h!]
%\centering
%\includegraphics[scale=1.0]{Figure5_color.pdf}
%\caption{Same as in Figure \ref{FG:HHPP_sPP} for a blend of 50:50 PIB/sPP.}
%\label{FG:PIB_sPP}
%\end{figure}

%\begin{figure}[h!]
%\centering
%\includegraphics[scale=1.0]{Figure6_color.pdf}
%\caption{Same as in Figure \ref{FG:HHPP_sPP} for a blend of 75:25 iPP/PE.}
%\label{FG:iPP_PE}
%\end{figure}

\subsection{Applications to Thermal Blends with Realistic $\chi$ Parameters}
In the multiscale modeling procedure presented above, we set $\chi=0$ in our soft-colloidal representation, and the UA simulations to which we combine our mesoscale simulations are assumed to be well-mixed. This provides an efficient means of obtaining pair correlation functions for polymer mixtures close to the athermal limit. In this section we examine the assumption that the blends are well-mixed  and discuss the application of our multiscale modeling procedure to modeling thermal mixtures where subtle local features such as monomer shape, branching, and site energetic asymmetries gives rise to non-trivial divergent fluctuations.

Blends of polypropylenes of different tacticities have been studied by Woo et al., and the $\chi$ parameter of blends of iPP/sPP and aPP/sPP was found to be nearly zero over the temperature range T=423-453 K.\cite{Woo} Thus, for the hhPP/sPP blend of Figure \ref{FG:HHPP_sPP} it is reasonable to assume the effective $\chi$ parameter will be small, which is supported by the overall quantitative agreement in Figure \ref{FG:MS} and \ref{FG:HHPP_PE}. For the other systems investigated in this publication, many of the pertinent $\chi$ parameters may be determined from the available literature on polyolefin blends which is presented in Table \ref{TB:Therm}. 

\begin{table}[h!b!p!] 
\caption{Temperature Dependence of Polyolefin Blends}
\begin{minipage}{3.5in}
\renewcommand{\thefootnote}{\alph{footnote}}
\begin{tabular}{cccc}
  \hline \hline
Blend [A/B] & $\chi(T)$ & $\chi/\chi_s$ ($453K$)  \\ 
\hline 
hhPP/PE & $-0.0294+17.58/T$\footnotemark[1] & 0.077 \\
PIB/PE & $0.00257+4.99/T$\footnotemark[2]  & 0.163 \\
iPP/PE & 0.01\footnotemark[3] & 0.360  \\
hhPP/PIB & $0.027 -11.4/T$\footnotemark[4] & 0.018 \\
\hline \hline
\end{tabular}
\end{minipage}
\label{TB:Therm}
$^{\text{a}}$ Reference 26
$^{\text{b}}$ Reference 27
$^{\text{c}}$ Reference 25
$^{\text{d}}$ Reference 23
\end{table}

To demonstrate the extension of our approach to thermal mixtures, we use the literature values for the $\chi$ parameter evaluated at the simulation temperature, $T=453K$ and perform MS-MD simulations for the mixtures: hhPP/PE, PIB/PE, and hhPP/PIB. The hhPP/PIB blend is particularly interesting because it exhibits a lower critical solution temperature (LCST) phase diagram, which can in principle be modeled using our multiscale approach as well.\cite{McCarty1} Since the parameters used in our mutiscale modeling procedure are determined using a UA description where each $\text{CH}_x$ group is represented as a effective site, care must be taken when calculating the effective parameter from Table \ref{TB:Therm} where $\chi$ is defined on a monomer basis, for which several $\text{CH}_x$ groups are grouped as one monomer unit. Since $\chi$ is proportional to the free energy of mixing per site, the $\chi$ parameter must be normalized by the average number of united atom sites per monomer\cite{JARAMILLO} as defined in the literature  (6 for hhPP/PE\cite{JEON}, 4 for PIB/PE\cite{LEE}, and 4.8 for hhPP/PIB\cite{JARAMILLO}). For hhPP/PE this corresponds to a value of $\chi = 0.0016$; for PIB/PE a value of $\chi = 0.0034$; and hhPP/PIB a value of $3.8\times 10^{-4}$. For the temperature range at which the united atom simulations were performed, the magnitude of $\chi$ in all cases is small, supporting our initial modeling of $\chi=0$ in the previous section. In Table \ref{TB:Therm} we present the thermodynamically relevant parameter, $\chi/\chi_s$ at $453K$, which provides an indication of how far the mixture is from the spinodal.

We calculated the total correlation function for the thermal mixtures of hhPP/PE, PIB/PE, and hhPP/PIB, using the multiscale modeling approach described above. These results are presented in the supplemental material\cite{Supplemental} and look nearly identical to the athermal results shown in Figures \ref{FG:HHPP_PE} and \ref{FG:PIB_PE}. Although the correlation functions are nearly identical with the athermal limit because the blends are well mixed, subtle differences in the structure as a result of increased concentration fluctuations should be manifest in the distributions. To assess the changes in structure that result from increased fluctuation, the monomer level partial static structure factors can be calculated from the total correlation function, $h^{mm}(k)$,
\begin{eqnarray}
S^{mm}_{AA}(k) &=& \phi \omega^{mm}_{AA}(k) + \rho \phi^2 h^{mm}_{AA}(k) \nonumber \ , \\
S^{mm}_{BB}(k) &=& (1- \phi)\omega^{mm}_{BB}(k) +\rho(1- \phi )^2 h^{mm}_{BB}(k) \ , \\
S^{mm}_{AB}(k) &=& \rho \phi (1-\phi ) h^{mm}_{AB}(k) \nonumber \ ,
\label{EQ:Smmk}
\end{eqnarray}
where the monomer form factors were determined from UA simulations as in Equation \ref{EQ:BLND10} above. The structure factor measuring correlations in the relative concentration, $S^{\phi \phi}(k)$, which diverges as the mixture approaches the spinodal, is expressed as a linear combination of these partial structure factors, 
\begin{equation}
S^{\phi \phi}(k)=(1-\phi)^2 S^{mm}_{AA}(k) + \phi^2 S^{mm}_{BB}(k) -2 \phi (1-\phi) S^{mm}_{AB}(k).
\label{EQ:Scck}
\end{equation}
In small angle neutron scattering (SANS) experiments, the $\chi$ parameter is determined from fitting the partial structure factor to the random phase approximation (RPA) equation of de Gennes\cite{deGennes}
\begin{equation}
\frac{1}{S(k)}=\frac{1}{\phi \omega^{mm}_{AA}(k)}+\frac{1}{(1-\phi)\omega^{mm}_{BB}(k)}-2\chi,
\label{EQ:SANS}
\end{equation}
where for convenience the monomer site volumes were set equal to one.

%\begin{figure}[h!]
%\includegraphics[scale=.8]{Figure7color.pdf}
%\caption{(A) The concentration fluctuation structure factor obtained from the multiscale modeling procedure (open circles) for hhPP/PE ($\phi=0.5$) at $T=453K$. For comparison the RPA equation evaluated at $\chi=0.0016$ (solid line) and $\chi=0.00$ (dashed line) is shown. Filled circles represent the structure factor from the full UA simulation. (B) The same as part (A), except for PIB/PE ($\phi=0.5$), for which the RPA equation, evaluated at $\chi=0.0034$ (solid line) and $\chi=0.00$ (dashed line), is shown. (C) the same except for the hhPP/PIB blend, and the RPA equation is evaluated at $\chi=0.00038$ (solid line) and $\chi=0.00$ (dashed line). }
%\label{FG:SANS}
%\end{figure}
\begin{figure}[h!]
\includegraphics[scale=.8]{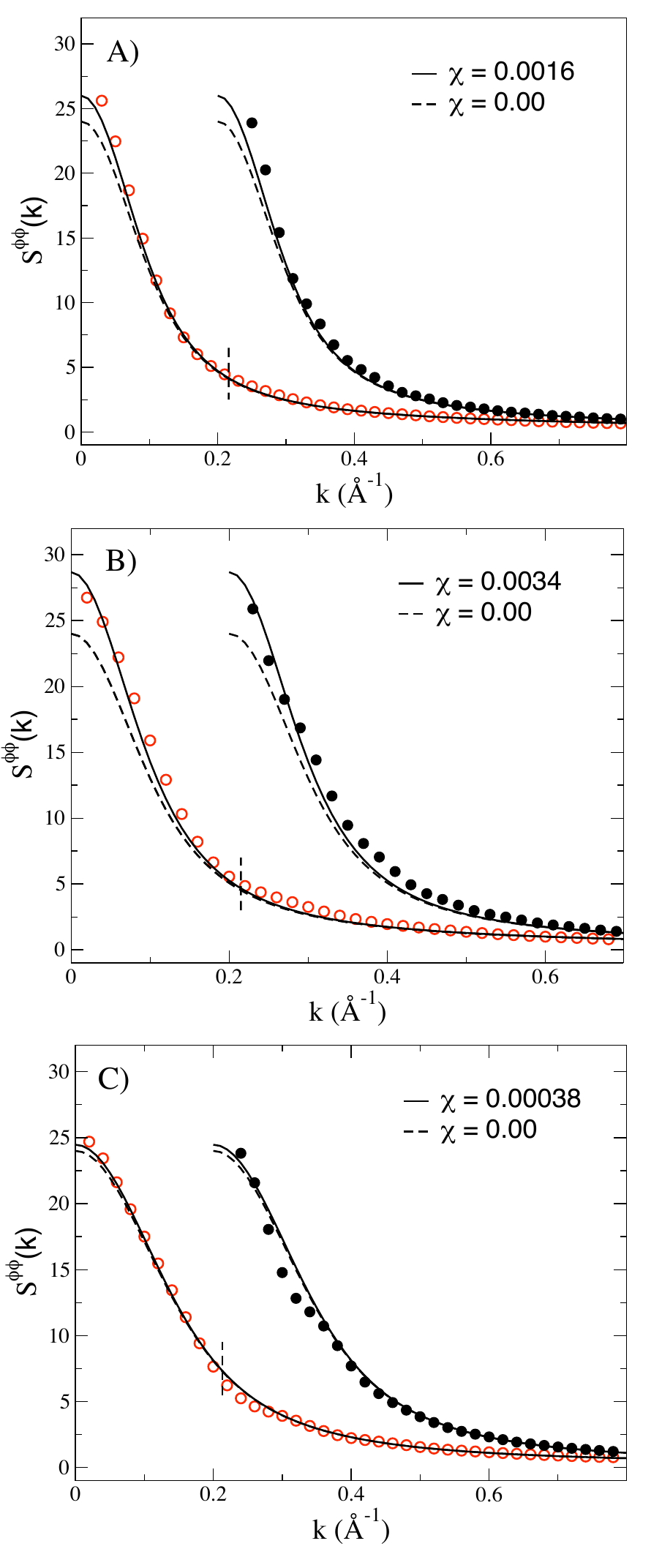}
\caption{(A) The concentration fluctuation structure factor obtained from the multiscale modeling procedure (open circles) for hhPP/PE ($\phi=0.5$) at $T=453K$. For comparison the RPA equation evaluated at $\chi=0.0016$ (solid line) and $\chi=0.00$ (dashed line) is shown. Filled circles represent the structure factor from the full UA simulation. (B) The same as part (A), except for PIB/PE ($\phi=0.5$), for which the RPA equation, evaluated at $\chi=0.0034$ (solid line) and $\chi=0.00$ (dashed line), is shown. (C) the same except for the hhPP/PIB blend, and the RPA equation is evaluated at $\chi=0.00038$ (solid line) and $\chi=0.00$ (dashed line). }
\label{FG:SANS}
\end{figure}
Figure \ref{FG:SANS} presents $S^{\phi \phi}(k)$ for the three different thermal blends obtained from our multiscale modeling procedure.  The static structure factor calculated in this manner exhibits good agreement when compared to the RPA equation, which was evaluated at $\chi=0.0016$ for hhPP/PE, $\chi=0.0034$ for PIB/PE, and $\chi = 3.8\times 10^{-4}$ for hhPP/PIB, using the intramolecular form factors from UA MD simulations.  The relevance of  this comparison to the RPA formula, which is known to fit low wavevector scattering curves well, is that there is clearly a better fit using the experimental $\chi$ parameter than with the $\chi=0$ case, which demonstrates the consistency of the description and that our approach is able to capture the fluctuations in concentration that arise in thermal polymer mixtures even at the relatively high temperatures of these simulations. Also, the advantage of a multiscale approach is exhibited by Figure \ref{FG:SANS} since the low wave vector regime is determined by mesoscale simulations so that UA simulations only need to be performed on systems at length scales up to the cut-off length (dashed line). This is important since only the initial stages of the divergent behavior need to be captured by united atom simulations, thus the need for prohibitively large simulation boxes is circumvented. For comparison, $S(k)$, calculated for the full UA MD simulation is also shown (offset for clarity) and agrees with our multiscale results, demonstrating the consistency between the two approaches.

When experimental values of the $\chi$ parameter are unavailable, a commonly used alternative is to estimate the interaction parameter from the solubility parameters calculated from a pure melt. These may be obtained from the cohesive energy determined from MD simulations of the individual melts. For the systems investigated here, we used united atom simulations of pure melts provided by Jaramillo et al.\cite{JARAMILLO} The cohesive energy density can be calculated from the radial distribution functions for each UA site type, $g_{ij}$, where $i,j$ designate a particular C, CH, CH$_2$, or CH$_3$ group, according to
\begin{equation}
U_{\text{coh}}=2 \pi \sum_{i j}\rho_{i} \rho_{j}  \int v_{i j}(r) g_{i j}(r) r^2 dr, 
\label{EQ:Coh}
\end{equation}
where $v_{ij}$ is the potential between nonbonded sites, for which we employ a Lennard-Jones potential with the corresponding TraPPE-UA parameters of Martin and Siepmann\cite{Siepmann}. In Equation \ref{EQ:Coh} the integration is carried out over the attractive branch of the potential. The solubility parameter ($\delta = \sqrt{-U_{\text{coh}}}$), calculated for each of the polyolefins in this study 
is presented in Table \ref{TB:Sol}. For comparison we also present results obtained by P\"utz, et al.\cite{Putz} from UA simulations of chains with 24 backbone carbons. 

\begin{table}[h!] 
\centering
\caption{Solubility Parameter for UA simulations of polyolefin melts (N=96) at $T=453K$. Values for $\delta_{MD-2}$ are computed from Ref. \cite{Putz} for chains with 24 backbone carbons.}
\begin{tabular}{ccc}
  \hline \hline
Polymer & $\delta$ (MPa)$^{\frac{1}{2}}$ & $\delta_{MD-2}$  \\
\hline 
PE & 14.4 & 13.7 ($T=448K$) \\
hhPP & 13.4 & 13.3   \\
iPP & 12.5 & 12.4  \\
PIB & 13.9 & 14.1  \\
sPP & 13.0 & 12.2 \\
\hline \hline
\label{TB:Sol}
\end{tabular}
\end{table}

From these solubility parameters, the Flory-Huggins parameter may be estimated according to 
\begin{equation}
\chi_H=\frac{(\delta_A - \delta_B)^2}{\sqrt{\rho^{\circ}_A \rho^{\circ}_B} k_B T}
\label{EQ:chisol}
\end{equation}
where the densities, $\rho^{\circ}$, are the pure melt densities of species A or B.  Table \ref{TB:chi} shows a comparison between the Flory-Huggins $\chi$ parameter determined from this conventional solubility parameters approach and the observed experimental values. A major limitation of this solubility approach is that  the value of $\chi$ is always positive and UCST phase behavior is always predicted. Furthermore, because non-combinatorial entropy is ignored, the concentration dependence of $\chi$ cannot be predicted. Despite these assumptions, the solubility approach has been demonstrated to hold up reasonably well for a large variety of polyolefin blends\cite{Lohse}. For comparison, in Table \ref{TB:chi} we also present the $\chi$ parameter computed directly from the full UA simulation of a polymer blend, by an extrapolation of the three lowest wave vector points to $k=0$ on an Ornstein-Zernike plot (1/S(k) vs. $k^2$), analogous to that obtained in typical SANS experiments. The lowest k vector is determined by the size of the simulation box and in this case is $k=0.4$ \AA. Due to this limitation, the effective $\chi$ parameter can be shifted by about $\pm 0.002$ while still maintaining a reasonable fit to the blend structure factor.\cite{HEINE}

\begin{table}[h!b!p!] 
\centering
\caption{Comparison between experiment, solubility parameter, and RPA approaches to determine the $\chi$ parameter ($T= 453K)$ for the polyolefin blends in Figure \ref{FG:SANS}}
\begin{tabular}{ccccccc}
\hline \hline
Blend [A/B] & $\chi_{\text{exp}}$ & $\chi_{H}$  &  $\chi_{\text{OZ plot}}$   \\
\hline
hhPP/PE & 0.0016 & 0.0048 &  0.0022 \\
PIB/PE & 0.0034 & 0.0012 &  0.0027 \\
hhPP/PIB & 0.00038 &  0.0012 & 0.0017  \\
\hline \hline
\label{TB:chi}
\end{tabular}
\end{table}

Finally we note that much work has been done using the PRISM theory of Schweizer and Curro to determine both the qualitative and quantitative miscibility behavior of polymer blends.\cite{Schweizer1994, Schweizer1993, Schweizer1997} For polyolefin blends of the type investigated here, realistic models utilizing PRISM expressions for the pair distribution function along with UA potential parameters have been shown to give reasonable estimates of $\chi$ for various polypropylene and polyethylene blends.\cite{Rajasekaran, Clancy, HEINE}; however this approach is not pursued further here since the purpose of the current publication is to demonstrate the implementation of the multiscale modeling procedure once the $\chi$ parameter is obtained.

\section{Conclusion and Outlook}
In this paper we present a multiscale modeling procedure to simulate mixtures of polymer chains at large scales. The method is completely general and applies to mixtures of polymers with different molecular structures, at different thermodynamic conditions of temperature, mixture composition, and chain length. 
The procedure combines large-scale information from a mesoscale simulation with a more detailed model which captures the local structure of the blend. For the detailed model, we use a united atom representation which captures the local polymer structure. Our mesoscale model represents the blend as a mixture of soft colloidal particles and accurately describes the large-scale structure as exhibited in the center of mass radial distribution function. Once the mesoscale simulations are performed, the relevant chemical details are reinserted by implementation of the Ornstein-Zernike formalism. We then combine this information with short united atom simulations to obtain a complete description over all length scales of interest. In this article we test the approach by applying our method to several different polyolefin mixtures. We check the validity of our procedure by direct comparison of the pair correlation function against full united atom molecular dynamic simulations. Because the UA-MD simulations available for comparison are limited to high temperatures where the samples can be more easily equilibrated, the test of our procedure includes only high temperature samples. However our mesoscale simulations are easily performed at temperatures approaching the demixing temperature. Furthermore, the multiscale modeling procedure provides a straightforward method to circumvent large simulations when modeling thermal mixtures, which has the potential to extend the range currently available to computer simulations of polymers. 
At low temperatures, where equilibration of even the local UA-MD simulation may be difficult, our two step procedure could be implemented ``on the fly" where information is transferred from the local UA-MD (which provides the packing information) to the global scale MS-MD and vice versa until consistency between the two is reached.

One advantage of the proposed method is that no optimized parameterization scheme is required to maintain consistency between the different levels of description. The approach is thus transferable to different systems and different conditions, while remaining quantitative enough to distinguish even subtle differences in chemical structure. Considering the advantages of the proposed multiscale approach, it should provide an important tool for investigating the dynamics of large-scale phenomena via MD simulation. Future work should focus on performing simulations of large systems approaching the spinodal where full UA MD simulations are prohibitively difficult.

\section{Acknowledgements}
We acknowledge support from the National Science Foundation through grant DMR-0804145.
Computational resources were provided by LONI through the TeraGrid project supported by NSF.

\newpage

\end{document}